# ORIENTATIONAL EFFECTS IN MIXTURES OF ORGANIC CARBONATES WITH ALKANES OR 1-ALKANOLS

JUAN ANTONIO GONZÁLEZ[(1)*], FERNANDO HEVIA[(1)], CRISTINA ALONSO-TRISTÁN[(2)], ISAÍAS GARCÍA DE LA FUENTE[(1)] AND JOSE CARLOS COBOS[(1)]

[(1)] G.E.T.E.F., Departamento de Física Aplicada, Facultad de Ciencias, Universidad de Valladolid, Paseo de Belén, 7, 47011 Valladolid, Spain,

*e-mail: jagl@termo.uva.es; Fax: +34-983-423136; Tel: +34-983-423757

[(2)] Dpto. Ingeniería Electromecánica. Escuela Politécnica Superior. Avda. Cantabria s/n. 09006 Burgos, (Spain)

**Abstract**

Interactions and structure of organic carbonate + alkane, and 1-alkanol + organic carbonate mixtures have been investigated by means of a set of molar excess functions, enthalpies ($H_m^E$), volumes ($V_m^E$), isobaric heat capacities, ($C_{pm}^E$) or entropies; and considering internal pressure ($P_{int}$); liquid-liquid equilibria or permittivity data. In addition, the mentioned systems have been studied using the Flory model and the concentration-concentration structure factor, $S_{CC}(0)$, formalism. The mixtures under consideration are characterized by dipolar interactions and by homocoordination (that is, by interactions between like molecules). In systems with a given solvent, dipolar interactions are weakened in the order: propylene carbonate (PC) > dimethyl carbonate (DMC) > diethyl carbonate (DEC). Comparison of mixtures containing DMC or DEC with those involving 2-propanone or 3-pentanone shows that dipolar interactions are not determined merely by values of the dipole moment, but they also depend on the size group. The enthalpies of the alkanol-carbonate interactions have been evaluated from calorimetric data. They are stronger in DMC solutions, and become weaker when the alcohol size increases in mixtures with a given carbonate. Application of the Flory model to 43 systems of the type 1-alkanol + carbonate provides a mean relative standard deviation for $H_m^E$ equal to 0.107. Results reveal that orientational effects decrease in the order DEC > PC > DMC. Orientational effects are particularly relevant in methanol or ethanol + DEC mixtures. Interestingly, the mentioned effects are weaker in 1-alkanol + DMC mixtures than in DMC + alkane systems. A similar trend is observed in DEC solutions when the considered alcohol is longer than ethanol.

Keywords: carbonates; thermodynamic properties; Flory, $S_{CC}(0)$, orientational effects, homocoordination

# 1. Introduction

We are engaged in a systematic study on orientational effects in liquid mixtures by means of the Flory model [1-5]. Using this approach, we have investigated systems such as 1-alkanol + linear or cyclic monoether [6], or + linear polyether [7], or + alkanone [8], or + nitrile [9] or 1-butanol + alkoxyethanol [10], or ether + alkane [11], + benzene, or + toluene [112], or + $CCl_4$ [13] and now extend these studies to dimethyl (DMC), diethyl (DEC) or propylene (PC) carbonate + alkane, or + 1-alkanol mixtures. Solutions including DMC or DEC have been previously treated [14-16] in terms of the DISQUAC [17] and ERAS [18] models, or using the Kirkwood-Buff integrals [19,20]. Similarly, in the framework of the UNIFAC (Dortmund version) [21] and Nitta-Chao [22] models, interaction parameters for the carbonate/alkane contacts, when dialkyl carbonates are involved, have been reported [23,24]. In the present work, systems with DMC or DEC are also investigated using the concentration-concentration structure factor formalism [25]. In fact, it is of high interest to link thermodynamic properties of liquid mixtures with local deviations from the bulk composition. At least, two procedures exist to investigate fluctuations in a binary mixture [25-27]. In the Kirkwood-Buff integrals formalism [19,20], fluctuations in the number of molecules of each component and the cross fluctuations are considered. A different alternative, based on the Bhatia-Thorton partial structure factors [28], is concerned with the study of fluctuations in the number of molecules regardless of the components, the fluctuations in the mole fraction and the cross fluctuations. This approach was generalized to link the asymptotic behaviour of the ordering potential to the interchange energy parameters in the semi-phenomenological theories of thermodynamic properties of liquid solutions [29-32]. We have applied this approach to mixtures involving pyridines [33] or to 1-alkanol + cyclic ether [34], or + alkanone [8] systems. From a theoretical point of view, the study of mixtures including carbonates is interesting for different reasons. Firstly, the investigation of solutions with linear organic carbonates is a previous step to the analysis of cyclic carbonates, ethylene or propylene carbonate, and of aromatic carbonates, methyl phenylcarbonate, e.g. The latter is particularly important when systems including aromatic heteroatoms are considered as then proximity effects between the phenyl and the polar groups may change considerably the molecular properties, and hence the interaction parameters when mixtures of these compounds are treated theoretically [35]. On the other hand, propylene carbonate is an aprotic solvent of very high dipole moment (5.36 D [36]) which is interesting to be studied in view of its local structure [37,38]. Secondly, carbonates are characterized by possessing the large functional group OCOO and it is important to investigate if the group size is relevant when describing the thermodynamic properties of liquid mixtures including the mentioned group. In fact, if the group is too large with respect to the average intermolecular distances, the interaction potential involved could be so complex that no theory can describe it conveniently, in such way that the thermodynamic properties are poorly described by means of the selected theory. From a practical point of view, it must be remarked that linear, cyclic or aromatic carbonates are widely employed in the industry. For example, they are used in the synthesis of organic compounds [39], as pharmaceuticals and agricultural chemicals, and as solvents for many synthetic and natural resins [40]. They are also very important in the Li battery technology [41,42]. DMC is used in the replacement of hazardous chemicals [43,44], as fuel additive or in the design of new refrigerants [24,45]. DMC and methyl phenyl carbonate are important intermediates obtained in the production of polycarbonates from diphenyl carbonate and bisphenol following green procedures which do not involve the highly toxic phosgene process [46]. All this supports that we have largely contributed to the development of databases of carbonate mixtures reporting experimental data on vapor-liquid [47-50], liquid-liquid and solid-liquid equilibria [15,51-53], excess molar volumes ($V_m^E$) [54,55] and excess molar enthalpies ($H_m^E$) [56,57] for such type of systems.

## 2. Theories

### 2.1.1 Flory model

We present here a brief summary of the main equations and hypotheses of the theory [1-5]. (i) Molecules are divided into segments (arbitrarily chosen isomeric portions of a molecule). (ii) The mean intermolecular energy per contact is assumed to be proportional to $-\eta/v_s$ (where $\eta$ is a positive constant which characterizes the energy of interaction for a pair of neighbouring sites and $v_s$ is the segment volume). (iii) The configurational partition function is obtained assuming that the number of external degrees of freedom of the segments is lower than 3. In this way, restrictions on the precise location of a given segment by its neighbours in the same chain are taken into account. (iv) Random mixing is assumed. The probability of having species of kind $i$ neighbours to any given site is equal to the site fraction ($\theta_i$). In the case of very large total number of contact sites, the probability of formation of an interaction between contacts sites belonging to different liquids is $\theta_1\theta_2$. Under these hypotheses, the Flory equation of state is:

$$\frac{\bar{P}\bar{V}}{\bar{T}} = \frac{\bar{V}^{1/3}}{\bar{V}^{1/3}-1} - \frac{1}{\bar{V}\bar{T}} \qquad (1)$$

where $\bar{V} = V/V^*$; $\bar{P} = P/P^*$ and $\bar{T} = T/T^*$ are the reduced volume, pressure and temperature, respectively. Equation (1) is valid for pure liquids and liquid mixtures. For pure liquids, the reduction parameters, $V_i^*$, $P_i^*$ and $T_i^*$ are obtained from densities, $\rho_i$, isobaric expansion coefficients, $\alpha_{Pi}$, and isothermal compressibilities, $\kappa_{Ti}$, data. The corresponding expressions for reduction parameters for mixtures are given elsewhere [6]. $H_m^E$ is determined from

$$H_m^E = \frac{x_1 V_1^* \theta_2 X_{12}}{\overline{V}} + x_1 V_1^* P_1^* (\frac{1}{\overline{V_1}} - \frac{1}{\overline{V}}) + x_2 V_2^* P_2^* (\frac{1}{\overline{V_2}} - \frac{1}{\overline{V}}) \qquad (2)$$

All the symbols have their usual meaning [6]. In this expression, the part which depends directly on $X_{12}$ is termed the interaction contribution to $H_m^E$. The remaining terms are the so-called equation of state contribution to $H_m^E$. The reduced volume of the mixture, $\overline{V}$, in equation (2) is obtained from the equation of state. Therefore, the molar excess volume can be also calculated:

$$V_m^E = (x_1 V_1^* + x_2 V_2^*)(\overline{V} - \varphi_1 \overline{V_1} - \varphi_2 \overline{V_2}) \qquad (3)$$

*2.1.2 Estimation of the Flory interaction parameter*

$X_{12}$ is determined from a $H_m^E$ measurement at given composition from [6-8]:

$$X_{12} = \frac{x_1 P_1^* V_1^* (1 - \frac{\overline{T_1}}{\overline{T}}) + x_2 P_2^* V_2^* (1 - \frac{\overline{T_2}}{\overline{T}})}{x_1 V_1^* \theta_2} \qquad (4)$$

For the application of this expression, we note that $\overline{VT}$ is a function of $H_m^E$:

$$H_m^E = \frac{x_1 P_1^* V_1^*}{\overline{V_1}} + \frac{x_2 P_2^* V_2^*}{\overline{V_2}} + \frac{1}{\overline{VT}} (x_1 P_1^* V_1^* \overline{T_1} + x_2 P_2^* V_2^* \overline{T_2}) \qquad (5)$$

and that from the equation of state, $\overline{V} = \overline{V}(\overline{T})$. More details have been given elsewhere [6-8]. Equation (5) is a generalization of that previously given to calculate $X_{12}$ from $H_m^E$ at $x_1 = 0.5$ [58]. Properties of organic carbonates at 298.15 K, molar volumes, $V_i$, $\alpha_{Pi}$, $\kappa_{Ti}$, and the corresponding reduction parameters, $P_i^*$ and $V_i^*$, needed for calculations are listed in Table 1. For 1-alkanols and alkanes, values have been taken from the literature [7,11]. At $T \neq 298.15$ K, the mentioned properties were estimated using the same equations as in previous applications for the temperature dependence of density, $\alpha_P$ and $\gamma$ (= $\alpha_P/\kappa_T$) [7,11]. $X_{12}$ values determined from experimental $H_m^E$ data at $x_1 = 0.5$ are collected in Table 2.

## 2.2 The concentration-concentration structure factor

Mixture structure can be investigated by means of the $S_{CC}(0)$ function [25,27,29,30,59]:

$$S_{CC}(0) = \frac{RT}{(\partial^2 G^M / \partial x_1^2)_{P,T}} = \frac{x_1 x_2}{D} \quad (6)$$

with

$$D = \frac{x_1 x_2}{RT}(\partial^2 G^M / \partial x_1^2)_{P,T} = 1 + \frac{x_1 x_2}{RT}\left(\frac{\partial^2 G_m^E}{\partial x_1^2}\right)_{P,T} \quad (7)$$

In equations (6) and (7), $G^M, G_m^E$ stand for the molar Gibbs energy of mixing and the molar excess Gibbs energy, respectively. $D$ is a function closely related to thermodynamic stability [60-62]. For ideal mixtures, $G_m^{E,id} = 0$ (excess Gibbs energy of the ideal mixture); $D^{id} = 1$ and $S_{CC}(0) = x_1 x_2$. From stability conditions, $S_{CC}(0) > 0$. If a system is close to phase separation, $S_{CC}(0)$ must be large and positive ($\infty$, if the mixture presents a miscibility gap). In the case of compound formation between components, $S_{CC}(0)$ must be very low (0, in the limit). Therefore, $S_{CC}(0) > x_1 x_2$ ($D < 1$) indicates that the dominant trend in the system is the homocoordination (separation of the components), and the mixture is then less stable than the ideal. If $0 < S_{CC}(0) < x_1 x_2 = S_{CC}(0)^{id}$, ($D > 1$), the fluctuations in the system have been removed, and the dominant trend in the solution is heterocoordination (compound formation). In such a case, the system is more stable than ideal. Therefore, $S_{CC}(0)$ is an useful magnitude to evaluate the non-randomness in the mixture [27,59].

## 3. Results

Results on $H_m^E$ obtained from the Flory model using $X_{12}$ values at $x_1 = 0.5$ are listed in Table 2, which also contains the interactional contribution to $H_m^E$ at equimolar composition. Experimental and theoretical values for $H_m^E$ are compared graphically in Figures. 1-5. For clarity, Table 2 also includes the relative standard deviations for $H_m^E$ defined as:

$$\sigma_r(H_m^E) = \left[\frac{1}{N}\sum\left(\frac{H_{m,exp}^E - H_{m,calc}^E}{H_{m,exp}^E}\right)^2\right]^{1/2} \quad (8)$$

where $N$ (=19) is the number of data points, and $H_{m,\exp}^E$ stands for the smoothed $H_m^E$ values calculated at $\Delta x_1 = 0.05$ in the composition range [0.05,0.95] from polynomial expansions, previously checked, given in the original works. Table 3 lists the results obtained for the $S_{CC}(0)$ function (Figures 6,7), with $D$ values calculated from $G_m^E$ functions obtained using DISQUAC and the needed parameters previously reported. [14,15].

## 5.    Discussion

Below, we are referring to thermodynamic properties at equimolar composition and 298.15 K. On the other hand, $n$ and $n_{OH}$ stand for the number of C atoms in the $n$-alkane or 1-alkanol, respectively.

### *5.1    Organic carbonate + alkane*
### *5.1.1    Mixtures with dialkyl carbonates*

These systems are characterized by rather strong dipolar interactions. The following features support such statement. (i) Large and positive values of $H_m^E (n = 7)/\text{J·mol}^{-1}$ = 1988 (DMC) [56]; 1328 (DEC) [57]. For $n$ = 10, $H_m^E/\text{J·mol}^{-1}$ = 2205 (DMC) [56]; 1536 (DEC) [57] (Figure 1). (ii) Relatively high upper critical solutions temperatures (UCST) in the case of DMC mixtures [51]: 297.62 K ($n$ = 12); 307.61 K ($n$ = 14); 316.21 K ($n$ = 16). (iii) Low values of excess heat capacities at constant pressure, $C_{pm}^E (n = 7)/\text{J·mol}^{-1}\cdot\text{K}^{-1}$ = 2.83 (DMC); 0.056 (DEC) [63]. The $C_{pm}^E$ curves are W-shaped [63,64]. This seems to be a typical feature of systems where non-random effects exist, which become more important when the mixture temperature is close to its UCST [65]. Consequently, the maximum of the $C_{pm}^E$ curves of DMC systems increases rapidly with $n$ [63,64]. (iv) Large and positive $TS_m^E(= H_m^E - G_m^E)$ values (Table 3; Figure 8). The curves shown in Figure 8 were calculated using the DISQUAC model with interaction parameters for the carbonate/aliphatic contacts determined previously [14]. (v) The Kirkwood-Buff integrals $G_{ii}$ (i =1,2) are also large and positive, while the $G_{12}$ integrals are negative [16]. In addition, the $G_{11}$ curves show a maximum [16], a trend usually encountered in systems where strong interactions occur between molecules of the same species. These features also reveal that dipolar interactions are stronger in DMC systems. Accordingly with the $H_m^E$ data, the corresponding $V_m^E$ values are also very large and change in line with $H_m^E$ when $n$ increases. Thus, $V_m^E$ (DMC)/ cm$^3$·mol$^{-1}$ = 1.158 ($n$ = 7); 1.442 ($n$ = 10) (Table 4) [54]. One can conclude that the main contribution to $V_m^E$ arises from interactional effects. This is also supported by the strong positive dependence of $V_m^E$ with $T$. For the DEC + hexane mixture [66], $\frac{\Delta V_m^E}{\Delta T} = 0.014$ cm$^3 \cdot$mol$^{-1} \cdot$K$^{-1}$. Systems where structural effects are very important, hexane + hexadecane, e.g., are characterized by large negative $\frac{\Delta V_m^E}{\Delta T}$ values ($-0.013$ cm$^3 \cdot$mol$^{-1} \cdot$K$^{-1}$ for the mentioned solution) [67].

*5.1.2 Propylene carbonate systems*

There is a rather large database including LLE measurements for multicomponent mixtures including PC due, e.g., to its applications to extract aromatic hydrocarbons from napththa reformate [68-70]. In contrast, the corresponding data for binary systems with alkanes is scarce. For the PC + methylcyclohexane system, the liquid-liquid equilibrium temperature is, at $x_1$(PC) = 0.0431, 348.15 K [69], the decane mixture shows a miscibility gap at 403.15 K between [0.033, 0.953] in mole fraction of PC [71]; and the upper critical solution temperature for the 1-octene mixture is 423.15 K [72]. On the other hand, activity coefficients of alkanes in solvent PC are extremely large: 81.7 for the octane system at 303.15 K [73]. These results are consistent with the very large dipole moment of PC (see above) and, together with those shown in the previous subsection for DMC or DEC mixtures, allow conclude that dipolar interactions between carbonate molecules become stronger in the sequence: DEC < DMC < PC.

*5.2. 1-alkanol + organic carbonate*

*5.2.1 Enthalpies of the hydroxyl-carbonate interactions*

Neglecting structural effects [60,74], $H_m^E$ can be considered as the result of three contributions. The positive ones, $\Delta H_{OH-OH}$, $\Delta H_{CO3-CO3}$, come, respectively, from the breaking of alkanol-alkanol and carbonate-carbonate interactions upon mixing. The negative contribution, $\Delta H_{OH-CO3}$, is due to the new OH---OCOO interactions created along the mixing process. That is [75-78]:

$$H_m^E = \Delta H_{OH-OH} + \Delta H_{CO3-CO3} + \Delta H_{OH-CO3} \qquad (9)$$

The $\Delta H_{OH-CO3}$ term represents the enthalpy of the H-bonds between 1-alkanols and organic carbonates. An estimation of this magnitude can be conducted extending the equation (9) to $x_1 \to 0$ [78-80]. In such a case, $\Delta H_{OH-OH}$ and $\Delta H_{CO3-CO3}$ can be replaced by $H_{m1}^{E,\infty}$ (partial excess molar enthalpy at infinite dilution of the first component) of 1-alkanol or carbonate + heptane systems. Thus,

$$\Delta H_{OH-O} = H_{m1}^{E,\infty}(1-\text{alkanol} + \text{carbonate})$$

$$-H_{\text{m1}}^{\text{E},\infty}(1-\text{alkanol} + \text{heptane}) - H_{\text{m1}}^{\text{E},\infty}(\text{carbonate} + \text{heptane}) \qquad (10)$$

Certainly, this is a rough estimation of $\Delta H_{\text{OH-CO3}}$ values due to: i) $H_{\text{m1}}^{\text{E},\infty}$ data used were calculated from $H_{\text{m}}^{\text{E}}$ measurements over the entire mole fraction range. ii) For 1-alkanol + $n$-alkane systems, it was assumed that $H_{\text{m1}}^{\text{E},\infty}$ is independent of the alcohol, a common approach when applying association theories [18,81-83]. We have used in this work, as in previous applications [7,78], $H_{\text{m1}}^{\text{E},\infty} = 23.2$ kJ·mol$^{-1}$ [84-86]. Nevertheless, it should be remarked that the values of $\Delta H_{\text{OH-CO3}}$ collected in Table 5, for systems with DMC or DEC, are still meaningful as they were obtained following the same procedure that in other previous investigations, which allows to compare enthalpies of interaction between 1-alkanols and different organic solvents. Inspection of Table 5 shows: (i) For a given carbonate, $\Delta H_{\text{OH-CO3}}$ increases more or less smoothly with $n_{\text{OH}}$. (ii) For a given 1-alkanol, $\Delta H_{\text{OH-CO3}}$ is lower for mixtures with DMC. That is, the OH and OCOO groups are more sterically hindered in longer 1-akanols and DEC, respectively, and this leads to weaker interactions between unlike molecules in systems with such compounds. No data on $H_{\text{m1}}^{\text{E},\infty}$ for PC + alkane mixtures have been encountered in the literature, and, consequently, $\Delta H_{\text{OH-CO3}}$ values for 1-alkanol + PC systems remain unknown. Nevertheless, it is expected that their variation with $n_{\text{OH}}$ is similar to that observed for $\Delta H_{\text{OH-CO3}}$ in systems with DMC or DEC.

*5.2.2 Excess molar enthalpies*

The large and positive $H_{\text{m}}^{\text{E}}$ values of these systems (Table 2; Figures 2-5) indicate that the positive $\Delta H_{\text{OH-OH}}, \Delta H_{\text{CO3-CO3}}$ contributions are predominant over the negative $\Delta H_{\text{OH-CO3}}$ term. Except for methanol systems, $H_{\text{m}}^{\text{E}}$(1-alkanol + carbonate) > $H_{\text{m}}^{\text{E}}$ (1-alkanol + isomeric alkane). Thus, $H_{\text{m}}^{\text{E}}$($n_{\text{OH}}$ = 3)/J·mol$^{-1}$ = 1794 (DEC) [87] > 459 (pentane) [88]. Clearly, organic carbonates are good breakers of the alcohol self-association. A similar trend is also encountered for 1-alkanol + linear polyether [7], or + $n$-alkanone mixtures [8]. This result suggests that in the studied mixtures, dipolar interactions play an essential role. Consequently, the shape of the $H_{\text{m}}^{\text{E}}$ curves for 1-alkanol + $n$-alkane, or + organic carbonate mixtures greatly differ. Systems with $n$-alkanes are characterized by $H_{\text{m}}^{\text{E}}$ curves skewed towards lower mole fractions of 1-alkanol ($x_1$), as the self-association of this compound is more easily broken at such condition. Carbonate mixtures show much more symmetrical $H_{\text{m}}^{\text{E}}$ curves (Figures 2-5). Interestingly, the $H_{\text{m}}^{\text{E}}$ curve of methanol + DMC is skewed towards higher $x_1$ values, while that of the methanol

+ DEC is much more symmetrical (Figure 2). One can conclude then that dipolar interactions are more relevant in DMC mixtures. It should be mentioned the very strong dipolar interactions existing in 1-alkanol + ethylene carbonate (dipole moment 4.81 D [36]) systems as it is shown by the upper critical solution temperatures of these mixtures: 312 K (1-propanol); 345.5 (1-hexanol) [89].

Comparison of $H_m^E$ values of systems formed by a given dialkyl carbonate and 1-alkanol or alkane of similar size shows that $H_m^E$ is higher for the solution including 1-alkanol. Thus, $H_m^E$(DEC)/J·mol$^{-1}$ = 2159 (1-octanol) [87] > 1399 (octane) [57]. This indicates that interactions between unlike molecules are of low relevance in systems with longer 1-alkanols and that such compounds are good breakers of the dipolar interactions between carbonate molecules.

On the other hand, $H_m^E$ data available in the literature [90,91] for 1-butanol + DMC or + DEC mixtures reveal that $H_m^E$ is a function strongly dependent on temperature. For the mentioned systems, $\Delta H_m^E / \Delta T \approx$ 17 J·mol$^{-1}$·K$^{-1}$. Large positive $C_{pm}^E$/J·mol$^{-1}$·K$^{-1}$ values are typically encountered in mixtures characterized by alcohol self-association (12.6 for 1-propanol + heptane) [92]. In contrast, mixtures where dipolar interactions are relevant show low $C_{pm}^E$/ J·mol$^{-1}$·K$^{-1}$ values (0.96 for ethanol + DMF) [93]. The large $\Delta H_m^E / \Delta T$ values of these solutions may indicate that any type of interactions, including those between unlike molecules, is largely broken when $T$ is increased. Interestingly, for the methanol + PC mixture, $C_{pm}^E$ / J·mol$^{-1}$·K$^{-1}$ = 6.6 [94], and this newly remarks the relevance of dipolar interactions in PC systems.

*5.2.2.1 The effect of increasing $n_{OH}$ in systems with a given dialkyl carbonate*

At this condition, $H_m^E$ values increase (Table 2). This may be explained taking into account that both $\Delta H_{CO3-CO3}$ and $|\Delta H_{OH-CO3}|$ terms also increase with $n_{OH}$ (Table 5). The former is due to the larger aliphatic surface of longer 1-alkanols break a higher number of carbonate-carbonate interactions, as $H_m^E$ of linear organic carbonate + $n$-alkane increases with $n$ (Table 2). The latter is related to the weaker interactions which exist between long 1-alkanols and carbonates. The increased $\Delta H_{CO3-CO3}$ and $|\Delta H_{OH-CO3}|$ values are predominant over the expected decrease of the contribution from the breaking of alkanol-alkanol interactions. It should be remembered that $H_m^E$ of heptane systems increases from ethanol to 1-butanol and then slowly decreases [95]. Thus, 1-alkanol + *n*-alkane or + carbonate mixtures show a different variation of $H_m^E$ with $n_{OH}$, and this also supports the relevance of dipolar interactions in carbonate systems.

*5.2.2.2 The effect of replacing DMC by DEC in systems with a given 1-alkanol*

DMC mixtures are characterized by larger $H_m^E$ values than those containing DEC (see above, Table 2). This means that the positive difference ($\Delta H_{CO3\text{-}CO3}$(DMC) $- \Delta H_{CO3\text{-}CO3}$(DEC)) is predominant over the negative value of ($\Delta H_{OH\text{-}CO3}$(DMC) $- \Delta H_{OH\text{-}CO3}$(DEC)) and over the higher contribution from the breaking of alkanol-alkanol interactions by the larger aliphatic surface of DEC. Note that $H_m^E$ of mixtures formed by a given 1-alkanol and *n*-alkane, increases with *n* [95].

*5.2.2.3 The effect of replacing DMC by PC in systems with a given 1-alkanol*

The $H_m^E$ dependence with $n_{OH}$ for PC systems is similar to those observed for mixtures containing DMC or DEC, and may be explained in similar terms. The larger $H_m^E$(PC) values compared to those including DMC can be ascribed to the $H_m^E$ contribution from the breaking of the carbonate-carbonate interactions is higher in the case of PC solutions due to the large dipole moment of this carbonate; and reveal that dipolar interactions are more important in PC mixtures. The lower $H_m^E$ value of the 1-butanol + PC system (Table 2) does not fit within this picture, and should be taken with caution.

*5.2.3 Excess entropies*

An interesting study can be conducted in terms of the $TS_m^E$ magnitude (Table 3; Figure 8). As previously, the $TS_m^E$ curves of 1-alkanol + dialkyl carbonate mixtures were calculated using DISQUAC with interaction parameters available in the literature [14,15]. $TS_m^E$ values of 1-alkanol + *n*-alkane mixtures are negative over almost the entire composition range, which is related to the alcohol self-association [85]. Only at low mole fractions of alcohol, positive $TS_m^E$ values are encountered (Figure 8) [85], as interactions between alkanol molecules are then more easily broken by alkanes. In addition, $TS_m^E$ and $n_{OH}$ values increase in line, as the weaker self-association of longer 1-alkanols leads to increased values of $TS_m^E$. 1-Alkanol + DMC or + DEC mixtures show positive $TS_m^E$ values at any $x_1$ value (Table 3; Figure 8). This clearly underlines that association/solvation effects are of minor importance. Such effects become even weakened for longer 1-alkanols, as $TS_m^E$ increases with $n_{OH}$ (Table 3). Accordingly with the $H_m^E$ results, $TS_\text{m}^\text{E}$ values of mixtures with a given carbonate and 1-alkanol are higher than those of the corresponding system with an alkane of similar size. This is also supported by the $G_\text{ii}$ values determined for these solutions [16]. In fact, the $G_\text{ii}$ curves show maxima on both sides of the concentration range, which is characteristic of mixtures where interactions between like molecules exist. The observed decrease of $TS_\text{m}^\text{E}$ when DMC is replaced by DEC may be ascribed to the less polar nature of DEC and to a lower enthalpic contribution to $TS_\text{m}^\text{E}$.

*5.2.4    Excess molar volumes*

For many of the systems considered, $V_\text{m}^\text{E}$ values are positive (Table 4). Therefore, the main contribution to this excess function comes from interactional effects. Interestingly, for DEC mixtures, $H_\text{m}^\text{E}$/J·mol$^{-1}$ = 2248 (1-decanol) [87] > 1536 (decane) [57], while $V_\text{m}^\text{E}$ / cm$^3$·mol$^{-1}$ = 0.6386 (decanol) [96] < 1.0629 (decane) [55]. This clearly indicates that structural effects are also present in systems with 1-alkanols. The different sign of $H_\text{m}^\text{E}$ and $V_\text{m}^\text{E}$ for methanol + DMC ($V_\text{m}^\text{E}$ = − 0.0628 cm$^3$·mol$^{-1}$) [96] or + DEC ($V_\text{m}^\text{E}$ = − 0.048 cm$^3$·mol$^{-1}$) [87] mixtures, or for methanol (− 0.336 cm$^3$·mol$^{-1}$ [97]), or + 1-propanol − 0.037 cm$^3$·mol$^{-1}$ [97]) + PC mixtures supports our conclusion. On the other hand, both $H_\text{m}^\text{E}$ and $V_\text{m}^\text{E}$ magnitudes increase with $n_\text{OH}$, and the $V_\text{m}^\text{E}$ variation is closely related to that of the corresponding interactional contribution to this excess function.

*5.3    Excess molar internal energies at constant volume*

The large $H_\text{m}^\text{E}$ values are not entirely due to interactional effects, but also to structural effects. The former are more properly considered using $U_\text{Vm}^\text{E}$, the excess internal energy at constant volume. Neglecting terms of higher order in $V_\text{m}^\text{E}$, $U_\text{Vm}^\text{E}$ can be written as [60,74]:

$$U_\text{Vm}^\text{E} = H_\text{m}^\text{E} - \frac{\alpha_\text{p}}{\kappa_\text{T}} T V_\text{m}^\text{E} \qquad (11)$$

where $\frac{\alpha_p}{\kappa_T} T V_m^\text{E}$ is the so-called equation of state (eos) contribution to $H_\text{m}^\text{E}$, and $\alpha_\text{p}$ is the isobaric thermal expansion coefficient of the mixture. Calculations of $\alpha_p$ and $\kappa_T$ were conducted on the basis of experimental data available in the literature on densities and adiabatic compressibilities of the involved mixtures [63,64,66,96,98-100], or assuming ideal behaviour ($F = \Phi_1 F_1 + \Phi_2 F_2$; $F_i$ is the property of the pure compound $i$) when such data are not available. The latter is a reasonable approximation in view of the rather low values of the excess values of $\alpha_p$ and $\kappa_T$ available in the literature [63,64,66,96,98-100]. Values of $\frac{\alpha_p}{\kappa_T}TV_m^E$ and $U_{Vm}^E$ are listed in Table 4. The eos contribution is very large for alkane mixtures and $U_{Vm}^E$ changes more smoothly with $n$ than $H_m^E$ does. A similar trend is observed for mixtures with 1-alkanols, although the eos contribution is lower. It is to be noted that, for a given 1-alkanol, the difference $U_{Vm}^E$ (PC) $-$ $U_{Vm}^E$ (DMC) is larger than that between the corresponding $H_m^E$ values, which underlines that dipolar interactions are stronger in PC mixtures.

### 5.4 Internal pressures

Internal pressures, $P_{int}$, have been determined using the expression [101-104]:

$$P_{int} = \frac{\alpha_p T}{\kappa_T} - p \qquad (12)$$

Values obtained in this work for some systems, using $\alpha_p$ and $\kappa_T$ data available in the literature [63,64,66,96,98-100], are collected in Table 6. From these results, some general trends can be stated. (i) $P_{int}$ values for carbonate + alkane mixtures are lower than those of 1-alkanol + carbonate mixtures. This may be due to the existence of interactions between unlike molecules and of structural effects in the latter systems; (ii) It is known that the main contributions to $P_{int}$ arise from dispersion forces and weak dipole-dipole interactions [103]; therefore, it seems that such type of interactions in 1-alkanol + carbonate mixtures become weaker in the order: PC > DMC > DEC. On the other hand, $P_{int}$ values can be also obtained from the equation [102]:

$$P_{int}^{VDW} = \frac{RT}{x_1 v_{f1} + x_1 v_{f2} + V_m^E} - P \qquad (13)$$

In this expression, $v_{fi}(=RT/(p + P_{int,i}))$ is the free volume of component i [102]. Experimental $P_{int}$ results are compared with those of $P_{int}^{VDW}$ in Table 6. The average difference between these magnitudes is 6.1% and this demonstrates that the Van der Waals equation is hold in large extent for the current mixtures, as eq. (13) is derived from this equation of state [102].

### 5.5 Dielectric constants and Kirkwood's correlation factor

We pay attention here to the permittivity data, $\varepsilon_r$, of the methanol + ethylene carbonate (EC) system which, as far as we know, are the only available in the literature for this type of mixtures [105]. Interestingly, the excess permittivity ($\varepsilon_r^E = \varepsilon_r - \phi_1\varepsilon_{r1} - \phi_2\varepsilon_{r2}$) of the solution is

− 3.1, much lower than the results for methanol + *N,N*-dimethylformamide (2.57 [106,107]), or + *N,N*-dimethylacetamide (0.52 [107,108]) systems. These very different results are remarkable as the considered amides have large dipole moments (3.7 D, *N,N*-dimethylformamide; 3.68 D, *N,N*-dimethylacetamide [109]). The large negative $\varepsilon_r^E$ value of the EC mixture reveals that the predominant trend in the solution is the breaking of the alcohol network and of the dipolar interactions between EC molecules in such way that a decrease of the dipolar polarization is produced [108]. The opposite behaviour is encountered in the mentioned systems with amides, characterized by an increase of the total effective dipole moment of the mixtures. This is supported by the calculation of the Kirkwood's correlation factor, $g_K$, of methanol + EC system according to the equation [110-112]:

$$g_K = \frac{9k_B T V_m \varepsilon_0 (\varepsilon_r - \varepsilon_r^\infty)(2\varepsilon_r + \varepsilon_r^\infty)}{N_A \mu^2 \varepsilon_r (\varepsilon_r^\infty + 2)^2} \qquad (14)$$

where the symbols have the usual meaning [107]. Details of the calculation procedure are given elsewhere. In absence of experimental measurements on density and refractive indices, molar, these magnitude were considered as ideal [113]. Results (Figure 9) show that the mixture structure slowly changes in a wide concentration range as the addition of methanol to EC does not lead to cooperative effects which increase the total effective dipole moment of the mixture. The methanol + DMF mixture behave differently (Figure 9). Experimental work on this matter is currently undertaken.

*5.6    The role of the size group in dialkyl carbonates*

Finally, it is pertinent to conduct a comparison between experimental results for systems with *n*-alkanones (particularly, 2-propanone, 3-pentanone) or with DMC or DEC. Firstly, it is necessary to remark that both the dipolar moment ($\mu$) and the effective dipolar moment ($\bar{\mu}$) are higher for *n*-alkanones. The dipole moments of the mentioned ketones are [114]: 2.88 D (2-propanone) and 2.82 D (3-pentanone); for the carbonates [109], $\mu$/D = 0.94 (DMC); 0.90 (DEC). The effective dipole moment is a useful magnitude to evaluate the impact of polarity on bulk properties and is defined by [60,115,116]

$$\bar{\mu} = \left(\frac{\mu^2 N_A}{4\pi \varepsilon_0 k_B V_m T}\right)^{1/2} \qquad (15)$$

Results for $\bar{\mu}$ are: 1.28 (2-propanone); 1.05 (3-pentanone); 0.39 (DMC); 0.31 (DEC). However for systems with a given alkane, say dodecane, UCST/K = 286.2 (2-propanone) [117] < 307.61 (DMC) [51]. Similarly, $H_m^E$ and $U_{Vm}^E$ values are also lower for alkanone mixtures including alkanes (Table S1; supplementary material). All this allows conclude that dipolar interactions are stronger in mixtures involving dialkyl carbonates and that the size group plays an important role when evaluating such interactions which are not merely determined by $\mu$ values. The high UCST value of the acetic anhydride + heptane system (342.52 K) [118] is consistent with this picture as, for acetic anhydride [109], $\mu$/D = 3.10 and $\bar{\mu}$ = 1.22. Regarding mixtures with a given 1-alkanol and 2-propanone, or 3-pentanone, they show lower $H_m^E$ values than those of the corresponding systems with DMC or DEC (Table S1, supplementary material). Thus, $H_m^E(n_{OH}=1)$/ J·mol$^{-1}$ = 686 (2-propanone) [119]; 1308 (DMC) [120]; 725 (3-pentanone) [121]; 1257 (DEC) [87]. This can be explained taking into account that the contribution to $H_m^E$ from the breaking of interactions between carbonate molecules is larger than that corresponding to the disruption of the ketone-ketone interactions. Moreover, interactions between unlike molecules are stronger in systems involving alkanones [8]. For example, the values $\Delta H_{OH-CO}(n_{OH}=1)$/kJ·mol$^{-1}$ = $-$28.6 (2-propanone); $-$24.2 (3-pentanone) are lower than those listed in Table 5 for methanol + DMC, or + DEC systems.

*5.7    Results from the Flory model*

The large and positive $X_{12}$ values obtained for the studied mixtures (Table 2) reveal that the main contribution to $H_m^E$ arises from interactions between like molecules. In the case of dialkyl carbonate mixtures, for enough large $n$ and $n_{OH}$ values, we note that $X_{12}(n) < X_{12}(n_{OH})$ ($n=n_{OH}$), in agreement with the trend encountered for the corresponding $H_m^E$ values. Thus, $X_{12}$(DEC)/J·cm$^{-3}$ = 73.08 (1-octanol) > 51.40 (octane) and $H_m^E$/J·mol$^{-1}$ = 2159 (1-octanol) [87] > 1399 (octane) [57]. That is, interactions between unlike molecules are of minor importance for systems with longer 1-alkanols. On the other hand, $X_{12}$ of DEC + $n$-alkane mixtures increases linearly with $n$: $X_{12}$ = 41.22 + 1.317$n$ ($r$ = 0.989). DMC systems behave somewhat differently as at 298.15 K, the temperatures of solutions with the longer $n$-alkanes are close to the UCST and the $H_m^E$ curves become flattened. Interestingly, $X_{12}$ and $H_m^E$ do not change in line with $n_{OH}$, as the former magnitude decreases when $n_{OH}$ is increased. This is due to the $H_{m,int}^E$ depends on $\theta_2 V_1^*$ (eq. (2)), a magnitude which is ranged for DEC solutions between 21.31 ($n_{OH}$ = 1) and 65.33 ($n_{OH}$ =10) cm$^3$·mol$^{-1}$.

Let's define the mean standard relative deviation of $H_m^E$ as:

$$\bar{\sigma}_{\text{r}}(H_{\text{m}}^{\text{E}}) = \frac{1}{N_{\text{S}}} \sum \sigma_{\text{r}}(H_{\text{m}}^{\text{E}}) \qquad (16)$$

where $N_{\text{S}}$ represents the number of systems considered. From the theoretical results using the Flory model, we can provide the following statements. (i) Orientational effects in 1-alkanol mixtures become stronger in the order: DMC < PC < DEC, as $\bar{\sigma}_{\text{r}}(H_{\text{m}}^{\text{E}}) = 0.086$ (DMC) < 0.109 (PC) < 0.132 (DEC). Orientational effects become weakened when $n_{\text{OH}}$ is increased, and are particularly relevant in DEC solutions with $n_{\text{OH}} = 1,2$ ($\bar{\sigma}_{\text{r}}(H_{\text{m}}^{\text{E}}) = 0.255$). (iii) For systems with DMC, orientational effects are weaker in solutions with 1-alkanols than in those with alkanes ($\bar{\sigma}_{\text{r}}(H_{\text{m}}^{\text{E}}) = 0.101$). This supports our previous statement about the ability of DMC as a breaker of the alcohol self-association. In addition, note that alkane systems are close to the UCST. (iv) The poor results obtained for methanol or ethanol + DEC mixtures remark that the alcohol self-association and/or interactions between unlike molecules are relatively more important in these mixtures. Orientational effects are also weaker in DEC systems with $n_{\text{OH}} > 3$ ($\bar{\sigma}_{\text{r}}(H_{\text{m}}^{\text{E}}) = 0.088$) than in those containing alkanes ($\bar{\sigma}_{\text{r}}(H_{\text{m}}^{\text{E}}) = 0.113$). (v) As usually, the increase of temperature leads to improve Flory results. It is interesting to compare experimental $H_{\text{m}}^{\text{E}}$ results at $T \neq 298.15$ K with results provided by the model using $X_{12}$ values determined from $H_{\text{m}}^{\text{E}}$ data at 298.15 K. Of course, theoretical results become then poorer as it is indicated by the following $H_{\text{m}}^{\text{E}}$/J·mol$^{-1}$ values for DMC systems: 1321 (methanol, $T = 313.15$ K); 1728 (ethanol, $T = 313.15$ K); 1990 (1-propanol, $T = 313.15$ K); 2104 (heptane, $T = 413.15$ K). The differences with experimental results (see Table 2), in the same order, are: $-14\%$; $-13.2\%$; $-15\%$ and 5.8%. We note that these differences are larger (and negative) for 1-alkanol systems, which clearly indicates that non-random effects are more relevant in such solutions. In fact, orientational effects show a stronger temperature dependence than effects related to dispersive interactions. It should be kept in mind that dipole-dipole interactions, or molecular anisotropy, are usually approximated by a spherical pair interaction inversely proportional to temperature [60,74]. (vi). The model provides very large $V_{\text{m}}^{\text{E}}$ values (Table 3) as the interactional contribution to this excess function is overestimated. Nevertheless, the variation $V_{\text{m}}^{\text{E}}$ with $n$ or $n_{\text{OH}}$ is correctly represented. The $V_{\text{m}}^{\text{E}}$ results can be better examined by means of the Prigogine-Flory-Patterson model (PFP) [122], where $V_{\text{m}}^{\text{E}}$ is written as the sum of three contributions: an interactional contribution, a curvature term and the so-called $P^*$ term. The second one depends on $-(\bar{V}_1 - \bar{V}_2)^2$ and is always negative. The latter depends on $(P_1^* - P_2^*)(\bar{V}_1 - \bar{V}_2)$. For the 1-alkanol + DEC systems considered, $P_1^* < P_2^*$; $\bar{V}_1 < \bar{V}_2$ and the $P^*$ term is always positive and increases rapidly with $n_{OH}$. This leads to large positive $V_m^E$ values as the curvature term is much lower than the $P^*$ contribution and the interactional contribution is also very large (Figure 10; Table S2, supplementary material). Similar trends are also valid for 1-alkanol + DMC, or + PC systems, although in the case of PC solutions, $P_1^* < P_2^*$; $\bar{V}_1 > \bar{V}_2$ and the $P^*$ contribution is negative.

We have investigated previously, using the Flory model, orientational effects in systems of the type: 1-alkanol + liner mono- [6] or polyether [7], + alkanone [8], or + alkanenitrile [9]. The $\bar{\sigma}(H_m^E)$ values change in the sequence: 0.323 (linear monoether) > 0.137 (linear polyether) > 0.114 (alkanenitrile) ≈ 0.107 (carbonate) ≈ 0.099 (n-alkanone). Clearly, orientational effects are stronger in systems with linear monoethers, where the alcohol self-association plays the main role. Results for linear polyether mixtures are improved when systems with methanol or ethanol are discarded ($\bar{\sigma}(H_m^E)$ = 0.054). Mixtures with alkanenitriles, n-alkanone or carbonates behave similarly.

### 5.8 $S_{CC}(0)$ results

Firstly, we note the very large $S_{CC}(0)$ values of ethanol or DMC + heptane mixtures, indicating that interactions between like molecules are predominant (Table 3; Figures 6-7). In the case of the ethanol system, it is due to the strong alcohol self-association. For all the examined systems, $S_{CC}(0) > 0.25$, and they are characterized by homocoordination. The large $S_{CC}(0)$ values of the DMC + heptane mixture can be ascribed to the proximity of the UCST at 298.15 K. Accordingly, the $S_{CC}(0)$ and LLE curves are skewed to higher carbonate concentrations. On the other hand, $S_{CC}(0)$ values of DMC systems are higher than those of DEC mixtures. That is, interactions between like molecules are more relevant in DMC solutions. When comparing $S_{CC}(0)$ results for 1-alkanol + DMC or DMC + heptane systems (Figure 6), we note that 1-alkanol systems show lower $S_{CC}(0)$ values, which can be ascribed to the new interactions between unlike molecules created upon mixing. In addition, $S_{CC}(0)$ increases with the alkanol size. One can conclude that alkanol-DMC interactions become then less relevant. This is supported by the variation of the symmetry of the $S_{CC}(0)$ curves. In fact, the mentioned symmetry is similar for the systems 1-hexanol + DMC and DMC + heptane as the curves are skewed to higher carbonate mole fractions. In contrast, the $S_{CC}(0)$ curve of the methanol solution is skewed to higher alkanol concentrations, which suggests that the self-association of this compound could be here more important (Figure 6). 1-Alkanol + DEC mixture behave differently (Figure 7). Thus, it is remarkable that $S_{CC}(0)$(methanol + DEC) > $S_{CC}(0)$(DEC + heptane), which may be ascribed to alkanol-alkanol interactions are relatively more relevant than those between unlike molecules. In constrast, for mixtures with enough long 1-alkanols, say 1-hexanol, $S_{CC}(0)$(1-hexanol + DEC) < $S_{CC}(0)$ (DEC + heptane), which points out that interactions between unlike molecules play now a more important role. On the other hand, $S_{CC}(0)$ increases up to 1-butanol and then decreases. This variation together with the change of the symmetry of the $S_{CC}(0)$ curves (Figure 7) might indicate that the role of the alkanol self-association becomes less relevant.

## 6. Conclusions

Organic carbonate + alkane, and 1-alkanol + organic carbonate mixtures have been investigated on the basis of different thermophysical properties and using the Flory model and the $S_{CC}(0)$ formalism. The studied systems are characterized by dipolar interactions and show homocoordination. For a given solvent, dipolar interactions become more relevant in the sequence: DEC < DMC < PC. It has been shown that dipolar interactions in systems with DMC or DEC are not determined merely by $\mu$ values, but they also depend on the size group. In mixtures with 1-alkanols, hydroxyl-carbonate mixtures are stronger in solutions with DMC, and become weaker when the alcohol size increases in systems with a given carbonate. Results from the Flory model show that orientational effects decrease in the order: DEC > PC > DEC. These effects seem to be particularly important in mixtures with methanol or ethanol. In systems containing DMC, orientational effects are weaker in 1-alkanol mixtures than in those containing alkanes. A similar trend is encountered in DEC systems for $n_{OH} > 2$.

**Funding**

The authors gratefully acknowledge the financial support received from the Consejería de Educación y Cultura of Junta de Castilla y León, under Project BU034U16. F. Hevia gratefully acknowledges the grant received from the program 'Ayudas para la Formación de Profesorado Universitario (convocatoria 2014), de los subprogramas de Formación y de Movilidad incluidos en el Programa Estatal de Promoción del Talento y su Empleabilidad, en el marco del Plan Estatal de Investigación Científica y Técnica y de Innovación 2013-2016, de la Secretaría de Estado de Educación, Formación Profesional y Universidades, Ministerio de Educación, Cultura y Deporte, Gobierno de España'.

## 7. List of symbols

| | |
|---|---|
| $C_p$ | heat capacity at constant pressure |
| $\Delta H$ | enthalpy of interaction |
| $g_K$ | Kirkwood's correlation factor (eq. 14) |
| $G$ | Gibbs energy |
| $H$ | enthalpy |
| $n$ | number of C atoms in *n*-alkane |
| $n_{OH}$ | number of C atoms in 1-alkanol |
| $P_{int}$ | internal pressure (eq. 12) |
| $P^*$ | reduction parameter for pressure in the Flory model |
| $S$ | entropy |
| $S_{CC}(0)$ | concentration-concentration structure factor (eq. 6) |
| $T$ | temperature |
| $U_V$ | internal energy at constant volume |
| $V$ | molar volume |
| $V^*$ | reduction parameter for volume in the Flory model |
| $x$ | mole fraction in liquid phase |

*Greek letters*

| | |
|---|---|
| $\alpha_P$ | isobaric thermal expansion coefficient |
| $\varepsilon_r$ | relative permittivity |
| $\kappa_T$ | isothermal compressibility |
| $\mu$ | dipole moment |
| $\bar{\mu}$ | effective dipole momento (eq. 15) |
| $\sigma_r$ | relative standard deviation (eq. 8) |
| $X_{12}$ | interaction parameter in the Flory model |

**Superscripts**

| | |
|---|---|
| E | excess property |

**Subscripts**

| | |
|---|---|
| i,j | compound in the mixture, (i, j =1,2) |
| m | molar property |

TABLE 1

Physical properties and Flory reduction parameters of organic carbonates at 298.15 K at pressure 0.1 MPa: $V_m$, molar volume; $\alpha_p$, isobaric coefficient of thermal expansion; $\kappa_T$, isothermal compressibility; $V_m^*$, reduction molar volume; and $P^*$, reduction pressure.

| Carbonate | $V_m$/cm$^3$·mol$^{-1}$ | $\alpha_p$/10$^{-3}$K$^{-1}$ | $\kappa_T$/TPa$^{-1}$ | $V_m^*$/cm$^3$·mol$^{-1}$ | $P^*$/MPa |
|---|---|---|---|---|---|
| dimethyl carbonate (DMC) [16] | 84.71 | 1.2541 | 893.1 | 65.28 | 704.8 |
| Diethyl carbonate (DEC) [16] | 121.90 | 1.2971 | 970 | 93.36 | 679.5 |
| Propylene carbonate (PC) [123] | 85.25 | 0.84 | 509 | 70.22 | 725 |

TABLE 2

Molar excess enthalpies, $H_m^E$, at equimolar composition and at temperature $T$ for dialkyl carbonate(1) + alkane(2) or 1-alkanol (1) + organic carbonate(2) systems. The interaction parameters, $X_{12}$, calculated from $H_m^E$ values at equimolar composition are also included.

| Compound | $T$/K | $H_m^E$ /J·mol⁻¹ | $H_{m,\text{int}}^E$ /J·mol⁻¹ | $X_{12}$ /J·cm⁻³ | $\sigma_r(H_m^E)^a$ | Ref. |
|---|---|---|---|---|---|---|
| Dimethyl carbonate(1) + Alkane (2) | | | | | | |
| Heptane | 298.15 | 1988 | 1418 | 96.93 | 0.112 | 56 |
| | 363.15 | 1972[b] | 1219 | 90.82 | 0.088 | 124 |
| | 413.15 | 1988[c] | 1079 | 85.87 | 0.065 | 124 |
| Octane | 298.15 | 2053 | 1485 | 97.47 | 0.117 | 56 |
| Decane | 298.15 | 2205 | 1631 | 100.57 | 0.112 | 56 |
| cyclohexane | 298.15 | 1947 | 1395 | 103.61 | 0.113 | 56 |
| Diethyl carbonate (1) + Alkane (2) | | | | | | |
| Hexane | 298.15 | 1264 | 895 | 49.74 | 0.137 | 57 |
| Heptane | 298.15 | 1328 | 951 | 50.26 | 0.111 | 57 |
| Octane | 298.15 | 1399 | 1015 | 51.40 | 0.098 | 57 |
| Decane | 298.15 | 1536 | 1142 | 54.01 | 0.095 | 57 |
| Tetradecane | 298.15 | 1798 | 1406 | 59.96 | 0.094 | 57 |
| Cyclohexane | 298.15 | 1320 | 949 | 55.19 | 0.142 | 57 |
| 1-Alkanol (1) + Dimethyl carbonate (2) | | | | | | |
| Methanol | 298.15 | 1308[d] | 936 | 125.84 | 0.086 | 120 |
| | 313.15 | 1542 | 1067 | 146.79 | 0.270 | 125 |
| | | 1457[d] | 1009 | 138.66 | 0.049 | 120 |
| | 328.15 | 1626[d] | 1086 | 152.76 | 0.046 | 120 |
| Ethanol | 298.15 | 1688[d] | 1220 | 124.08 | 0.076 | 120 |
| | 303.15 | 1805 | 1268 | 131.80 | 0.155 | 126 |
| | 313.15 | 1972 | 1374 | 142.95 | 0.163 | 125 |
| | | 1863[d] | 1300 | 135.11 | 0.047 | 120 |
| | 328.15 | 2069[d] | 1394 | 148.26 | 0.051 | 120 |
| 1-propanol | 298.15 | 1955[d] | 1428 | 120.34 | 0.073 | 120 |
| | 303.15 | 2156 | 1556 | 132.23 | 0.088 | 126 |
| | 313.15 | 2321 | 1638 | 141.39 | 0.092 | 125 |
| | | 2181[d] | 1543 | 132.95 | 0.046 | 120 |
| | 328.15 | 2372[d] | 1625 | 143.08 | 0.049 | 120 |

TABLE 2 (continued)

| | | | | | | |
|---|---|---|---|---|---|---|
| 1-butanol | 288.15 | 2126 | 1599 | 114.95 | 0.087 | 90 |
| | 298.15 | 2356 | 1737 | 126.63 | 0.082 | 90 |
| | 313.15 | 2570 | 1839 | 136.92 | 0.058 | 90 |
| 1-pentanol | 303.15 | 2469 | 1826 | 119.0 | 0.075 | 126 |
| 1-octanol | 303.15 | 2772 | 2110 | 106.6 | 0.050 | 126 |
| 1-Alkanol (1) + Diethyl carbonate (2) | | | | | | |
| Methanol | 298.15 | 1257 | 898 | 110.73 | 0.319 | 87 |
| Ethanol | 298.15 | 1523 | 1096 | 101.11 | 0.230 | 87 |
| | 303.15 | 1645 | 1171 | 108.81 | 0.217 | 127 |
| 1-propanol | 298.15 | 1794 | 1308 | 99.09 | 0.166 | 87 |
| | 303.15 | 1880 | 1357 | 103.54 | 0.147 | 127 |
| 1-butanol | 293.15 | 1858 | 1385 | 89.29 | 0.116 | 91 |
| | 298.15 | 1944 | 1435 | 93.19 | 0.134 | 87 |
| | 303.15 | 2076 | 1517 | 99.21 | 0.108 | 91 |
| | 313.15 | 2206 | 1581 | 104.83 | 0.091 | 91 |
| 1-pentanol | 298.15 | 1984 | 1485 | 85.10 | 0.070 | 128 |
| | 303.15 | 2036 | 1508 | 86.97 | 0.088 | 127 |
| 1-hexanol | 298.15 | 2016 | 1524 | 78.89 | 0.096 | 87 |
| 1-octanol | 298.15 | 2159 | 1666 | 73.08 | 0.072 | 87 |
| | 303.15 | 2277 | 1740 | 76.77 | 0.056 | 127 |
| 1-decanol | 298.15 | 2248 | 1757 | 67.93 | 0.063 | 87 |
| 1-Alkanol (1) + Propylene carbonate (2) | | | | | | |
| Methanol | 298.15 | 1362 | 1067 | 133.03 | 0.123 | 129 |
| | 323.15 | 1595 | 1205 | 154.38 | 0.196 | 129 |
| Ethanol | 298.15 | 1876 | 1457 | 137.96 | 0.103 | 130 |
| | | 1783 | 1387 | 131.17 | 0.091 | 129 |
| | 323.15 | 2102 | 1575 | 153.15 | 0.079 | 129 |
| 1-propanol | 298.15 | 2138 | 1662 | 130.69 | 0.126 | 130 |
| | | 2115 | 1644 | 129.29 | 0.084 | 129 |
| | 323.15 | 2455 | 1839 | 148.57 | 0.031 | 129 |
| 1-butanol | 298.15 | 2200 | 1715 | 116.66 | 0.153 | 130 |

[a] relative standard deviation (eq. 8) ; [b]$p = 1.584$ MPa ; [c]$p = 1.686$ MPa ; [d]$p = 1$ MPa

TABLE 3

Values of $TS_m^E (= H_m^E - G_m^E)$ and of concentration-concentration structure factor, $S_{CC}(0)$, calculated according the DISQUAC model for dialkyl carbonate (1) + $n$-alkane(2), or 1-alkanol(1) + dialkyl carbonate(2) mixtures at composition $x_1$, 298.15 K and 0.1 MPa.

| System | $TS_m^E (x_1 = 0.5) / J \cdot mol^{-1}$ | $x_1$ | $S_{CC}(0)_{max} (x_1)^a$ |
|---|---|---|---|
| DMC + heptane | 832 | 0.57 | 3.87 |
| DEC + heptane | 556 | 0.48 | 0.66 |
| Methanol + DMC | 176 | 0.59 | 0.83 |
| ethanol + DMC | 578 | 0.55 | 0.98 |
| 1-propanol + DMC | 937 | 0.51 | 1.17 |
| 1-butanol + DMC | 1211 | 0.47 | 1.56 |
| 1-hexanol + DMC | 1493 | 0.39 | 1.85 |
| Methanol + DEC | 303 | 0.59 | 0.72 |
| ethanol + DEC | 559 | 0.56 | 0.77 |
| 1-propanol + DEC | 622 | 052 | 0.89 |
| 1-butanol + DEC | 1002 | 0.50 | 0.74 |
| 1-hexanol + DEC | 1277 | 0.46 | 0.60 |
| 1-octanol+ DEC | 1408 | 0.43 | 0.55 |

$^a$maximum $S_{CC}(0)$ value

TABLE 4

Molar excess volumes, $V_m^E$, at 0.1 MPa, 298.15 K and equimolar composition for dialkyl carbonate(1) + n-alkane(2) and for 1-alkanol(1) + organic carbonate(2) systems. Comparison of experimental (exp.) results with Flory calculations using interaction parameters listed in Table 2. Also included are the equation of state contribution ($\frac{\alpha_p}{\kappa_T} TV_m^E$) to $H_m^E$ and the excess molar internal energy at constant volume, $U_{Vm}^E$.

| Compound | $V_m^E$ /cm$^3$·mol$^{-1}$ | | $\frac{\alpha_p}{\kappa_T} TV_m^E$ /J·mol$^{-1}$ | $U_{Vm}^E$ /J·mol$^{-1}$ | Ref. |
|---|---|---|---|---|---|
| | Exp. | Flory | | | |
| Dimethyl carbonate(1) + alkane (2) | | | | | |
| Heptane | 1.158 | 1.825 | 346 | 1642 | 54 |
| Decane | 1.4425 | 2.217 | 458 | 1747 | 54 |
| Diethyl carbonate(1) + alkane (2) | | | | | |
| Heptane | 0.7362 | 1.266 | 226 | 1102 | 55 |
| Octane | 0.8755 | 1.493 | 275 | 1124 | 55 |
| Decane | 1.0629 | 1.769 | 344 | 1192 | 55 |
| Tetradecane | 1.2403 | 2.147 | 416 | 1382 | 55 |
| 1-alkanol (1) + dimethyl carbonate (2) | | | | | |
| Methanol | −0.0628 | 1.048 | −23 | 1331 | 96 |
| Ethanol | 0.1505 | 1.492 | 53 | 1630 | 98 |
| 1-propanol | 0.3693 | 1.802 | 135 | 1823 | 98 |
| 1-butanol | 0.4771 | 2.192 | 191 | 2187 | 90 |
| 1-alkanol (1) + diethyl carbonate (2) | | | | | |
| Methanol | −0.048 | 1.054 | −23 | 1274 | 87 |
| Ethanol | 0.1143 | 1.405 | 39 | 1481 | 87 |
| 1-propanol | 0.2225 | 1.736 | 81 | 1714 | 87 |
| 1-butanol | 0.2815 | 1.931 | 98 | 1844 | 87 |
| 1-hexanol | 0.3940 | 2.032 | 141 | 1875 | 87 |
| 1-octanol | 0.5200 | 2.124 | 187 | 1973 | 87 |
| 1-decanol | 0.6386 | 2.120 | 232 | 2016 | 87 |
| 1-alkanol (1) + propylene carbonate (2) | | | | | |
| Methanol | −0.3362 | 0.352 | −128 | 1490 | 130 |
| Ethanol | −0.2021 | 0.631 | −85 | 1961 | 130 |
| 1-propanol | −0.037 | 0.931 | −14 | 2152 | 130 |

TABLE 4 (continued)

| | | | | | |
|---|---|---|---|---|---|
| 1-butanol | 0.067 | 1.057 | 25 | 2175 | 130 |

TABLE 5

Partial molar excess enthalpies,[a] $H_1^{E,\infty}$, at 298.15 K at 0.1 MPa for dialkyl carbonate(1) + oheptane(2) mixtures, and hydrogen bond enthalpies, $\Delta H_{\text{OH-CO3}}$, for 1-alkanol(1) + dialkyl carbonate(2) systems.

| System | $H_1^{E,\infty}$ /kJ·mol$^{-1}$ | $\Delta H_{\text{OH-CO3}}$ /kJ·mol$^{-1}$ |
|---|---|---|
| DMC(1) + heptane(2) | 9.54 [56] | |
| DEC(1)+ heptane(2) | 7.07 [57] | |
| Methanol(1) + DMC(2) | 5.78 [120] | −27.0 |
| ethanol(1) + DMC(2) | 6.98 [120] | −25.8 |
| 1-propanol(1) + DMC(2) | 7.38 [120] | −25.4 |
| 1-butanol(1) + DMC(2) | 13.22 [90] | −25.3 |
| Methanol(1) + DEC(2) | 8.04 [87] | −22.2 |
| ethanol(1) + DEC(2) | 9.53 [87] | −20.7 |
| 1-propanol(1) + DEC(2) | 10.35 [87] | −19.9 |
| 1-butanol(1) + DEC(2) | 11.28 [87] | −19.0 |
| 1-pentanol(1) + DEC(2) | 10.66 [128] | −19.6 |
| 1-hexanol(1) + DEC(2) | 12.24 [87] | −18.0 |
| 1-octanol(1) + DEC(2) | 12.42 [87] | −18.7 |
| 1-decanol(1) + DEC(2) | 14.19 [87] | −16.1 |

[a]values obtained from $H_m^E$ data over the whole concentration range

TABLE 6

Internal pressures of pure compounds, $P_{int,i}$, and for the mixtures dialkyl carbonate + $n$-alkane or 1-alkanol + carbonate, $P_{int}$ (eq. 12), at equimolar composition, 298.15 K and 0.1 MPa. Comparison with results obtained from the Van der Waals model, $P_{int}^{VDW}$ (eq. 13).

| System | $P_{int,1}$/MPa | $P_{int,2}$/MPa | $P_{int}$/MPa | $P_{int}^{VDW}$/MPa | Ref. |
|---|---|---|---|---|---|
| DMC(1) + $n$-C$_8$(2) | 418.6 | 264.5 | 302.3 | 276.5 | 64,131 |
| DMC(1) + $n$-C$_9$(2) | 418.6 | 273.9 | 308.6 | 279.0 | 64,132 |
| DEC(1) + $n$-C$_{10}$(2) | 398.6 | 267.2 | 306.8 | 289.7 | 133 |
| Methanol(1) + DMC(2) | 285.6 | 411.2 | 366.0 | 342.5 | 96,98 |
| Ethanol(1) + DMC(2) | 283.2 | 411.2 | 345.6 | 332.7 | 98 |
| 1-propanol(1) + DMC(2) | 291.7 | 411.2 | 363.4 | 321.7 | 98 |
| 1-butanol(1) + DMC(2) | 298.1 | 411.2 | 359.7 | 324.5 | 98 |
| Methanol(1) + DEC(2) | 285.6 | 356.4 | 335.8 | 335.8 | 99 |
| Ethanol(1) + DEC(2) | 283.2 | 356.4 | 340.0 | 326.2 | 99 |
| 1-propanol(1) + DEC(2) | 291.7 | 356.4 | 342.2 | 326.5 | 99 |
| 1-butanol(1) + DEC(2) | 298.1 | 356.4 | 335.1 | 328.0 | 99 |
| Ethanol(1) + PC(2) | 283.2 | 492 | 421.5 | 367.6 | 100,130 |
| 1-propanol(1) + PC(2) | 291.7 | 492 | 390.1 | 368.2 | 100,130 |
| 1-butanol(1) + PC(2) | 298.1 | 492 | 372.2 | 367.6 | 100,130 |

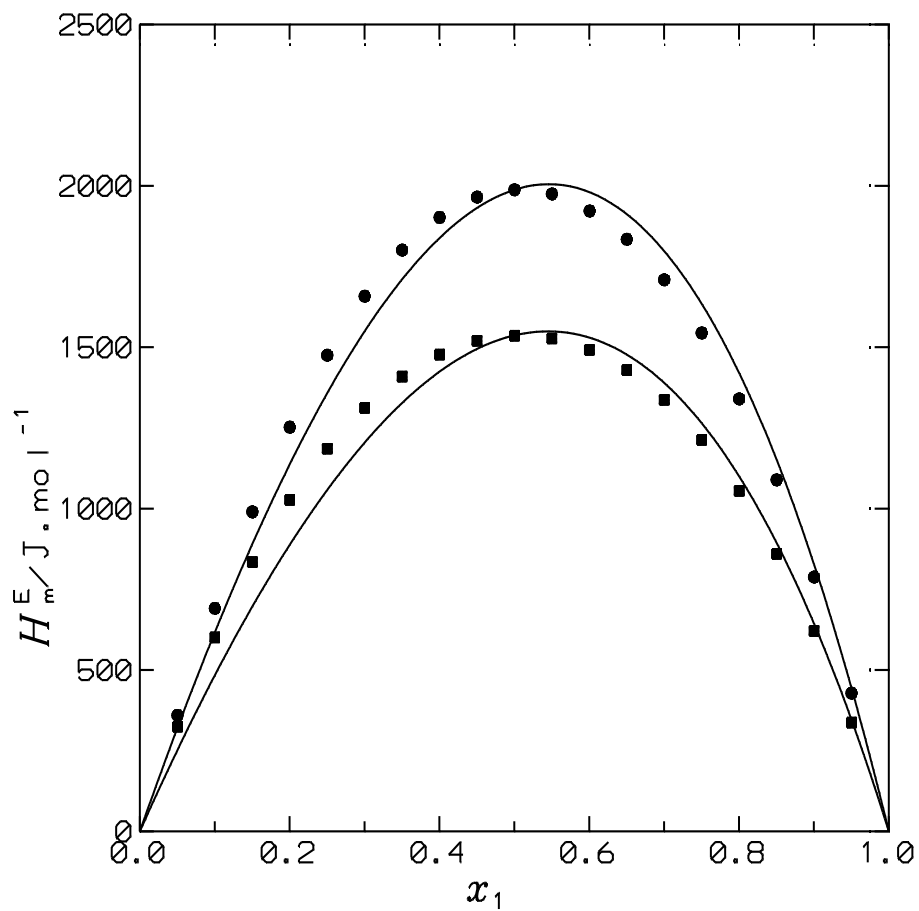

Figure 1. $H_m^E$ of dialkyl carbonate(1) + alkane(2) mixtures. Symbols, experimental results: (●), DMC(1) + heptane(2) ($T$ = 413.15 K; $P$ = 1.686 MPa) [124]; (■), DEC(1) + decane(2) ($T$ = 298.15 K) [57]. Solid lines, Flory calculations using interaction parameters listed in Table 2

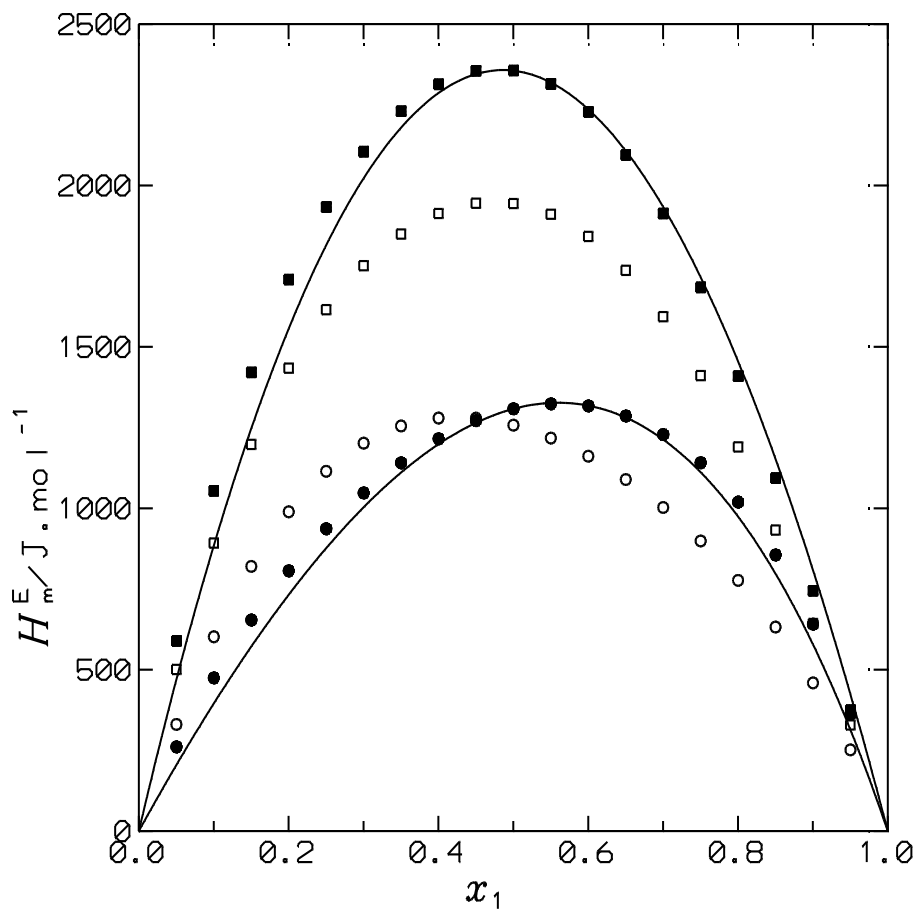

Figure 2. $H_m^E$ of 1-alkanol(1) + dialkyl carbonate(2) mixtures at 298.15 K. Symbols, experimental results. Full symbols, DMC mixtures [120] ($P$ = 1 MPa): (●), methanol; (■), 1-butanol. Open symbols, DEC systems [87] ($P$ = 0.1 MPa): (O), methanol; (□), 1-butanol. Solid lines, Flory calculations using interaction parameters listed in Table 2.

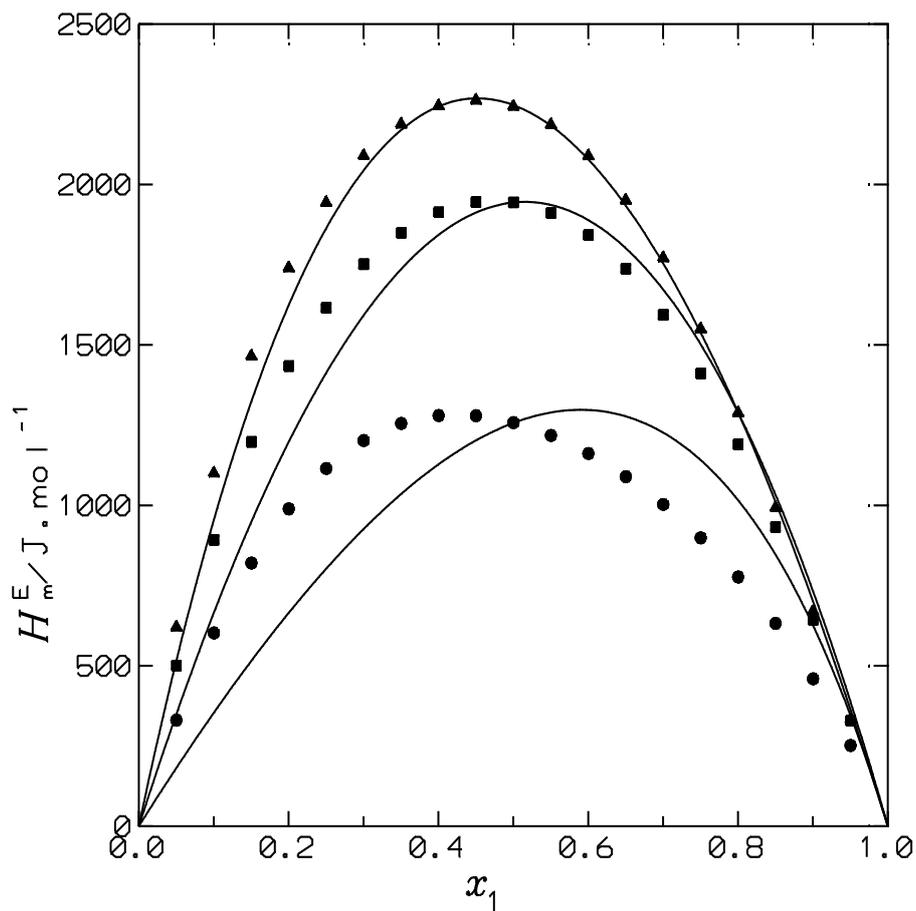

Figure 3. $H_m^E$ of 1-alkanol(1) + DEC(2) mixtures at 298.15 K and 0.1 MPa. Symbols, experimental results [87]: (●), methanol; (■), 1-butanol; (▲), 1-decanol. Solid lines, Flory calculations using interaction parameters listed in Table 2.

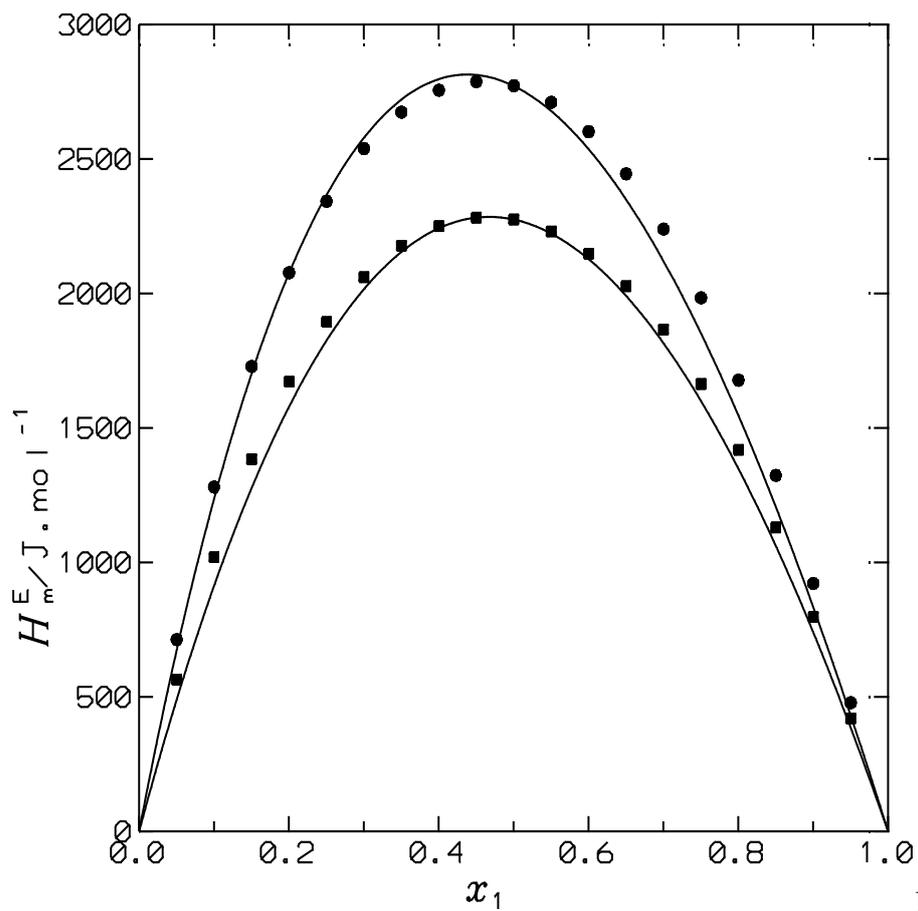

Figure 4. $H_\mathrm{m}^\mathrm{E}$ of 1-octanol(1) + DMC(2), or + DEC(2) mixtures at 303.15 K and 0.1 MPa. Symbols, experimental results: (●), DMC [126]; (■), DEC [127]. Solid lines, Flory calculations using interaction parameters listed in Table 2.

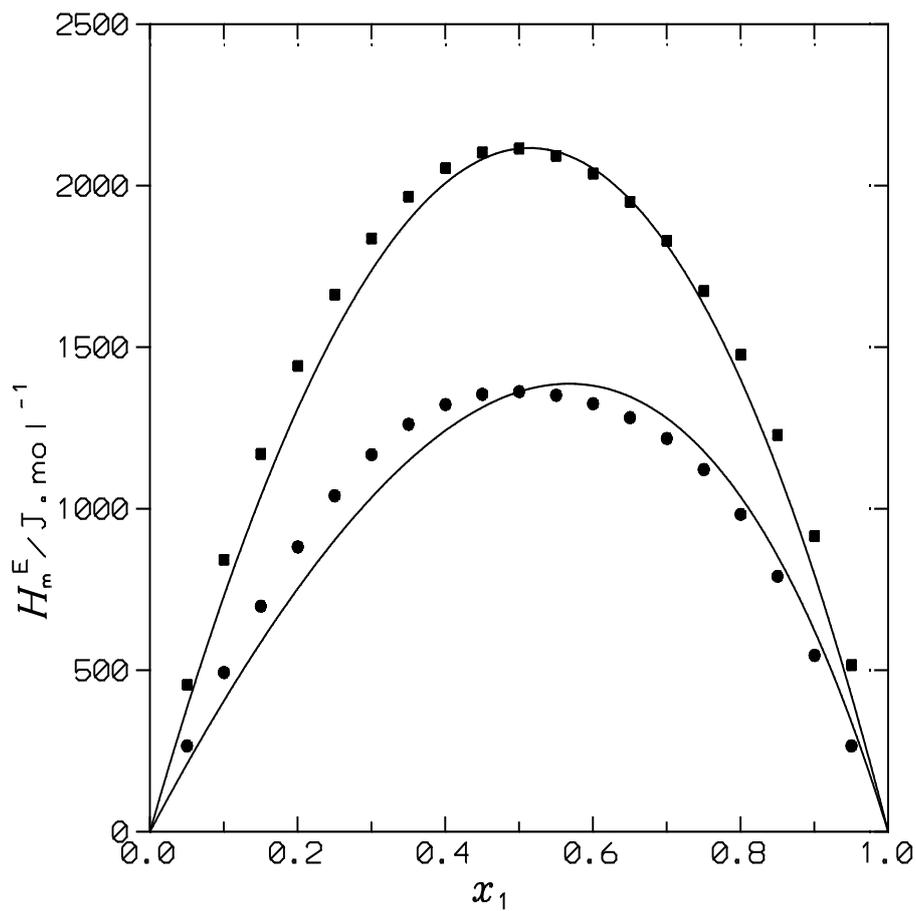

Figure 5. $H_\mathrm{m}^\mathrm{E}$ of methanol(1) or 1-propanol(1) + PC(2) mixtures at 298.15 K and 0.1 MPa. Symbols, experimental results [129]: (●), methanol; (■), 1-propanol Solid lines, Flory calculations using interaction parameters listed in Table 2

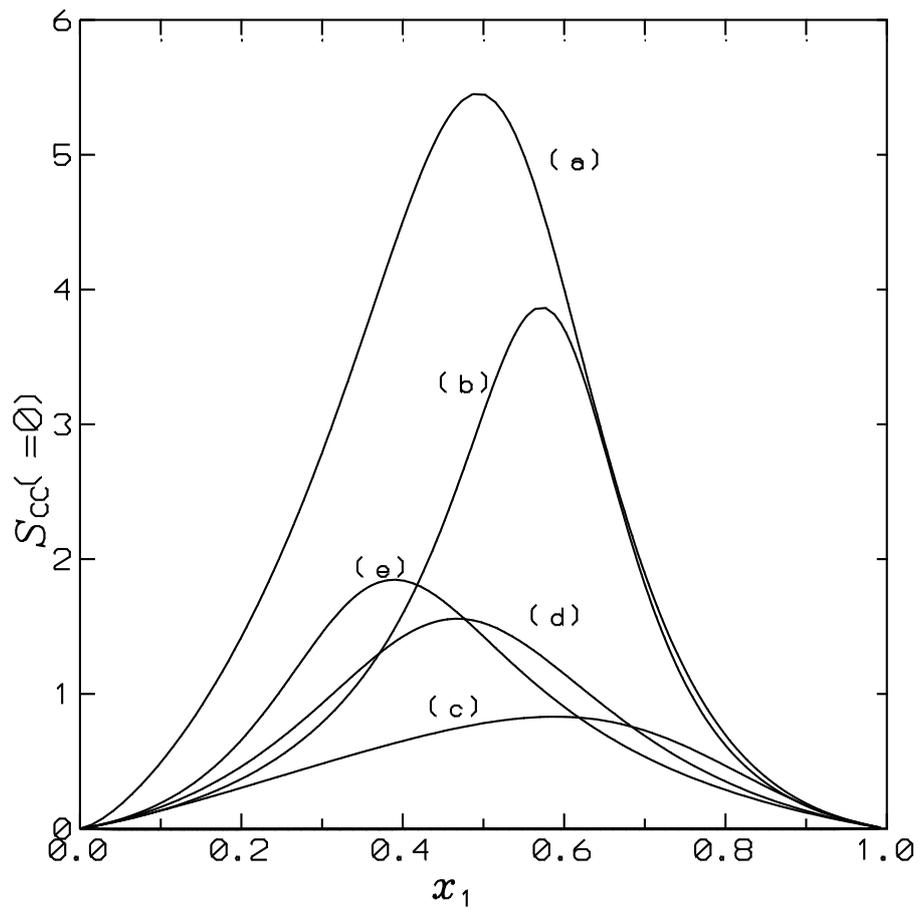

Figure 6. $S_{CC}(0)$ curves, obtained from the DISQUAC model, for systems at 298.15 K and 0.1 MPa: (a), ethanol(1) + heptane(2); (b), DMC(1) + heptane(2); (c), methanol(1) + DMC(2); (d), 1-butanol(1) + DMC(2); (e), 1-hexanol(1) + DMC(2).

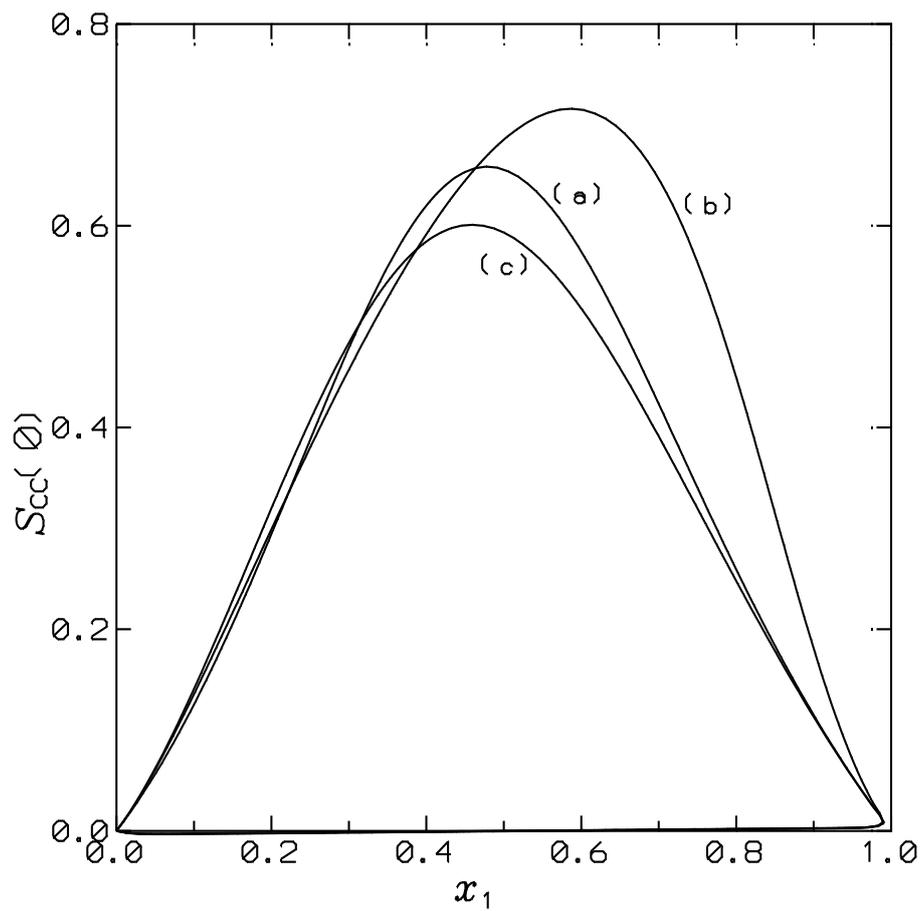

Figure 7. $S_{CC}(0)$ curves, obtained from the DISQUAC model, for systems at 298.15 K and 0.1 MPa: (a), DEC(1) + heptane(2); (a), methanol(1) + DEC(2); (c), 1-hexanol(1) + DEC(2).

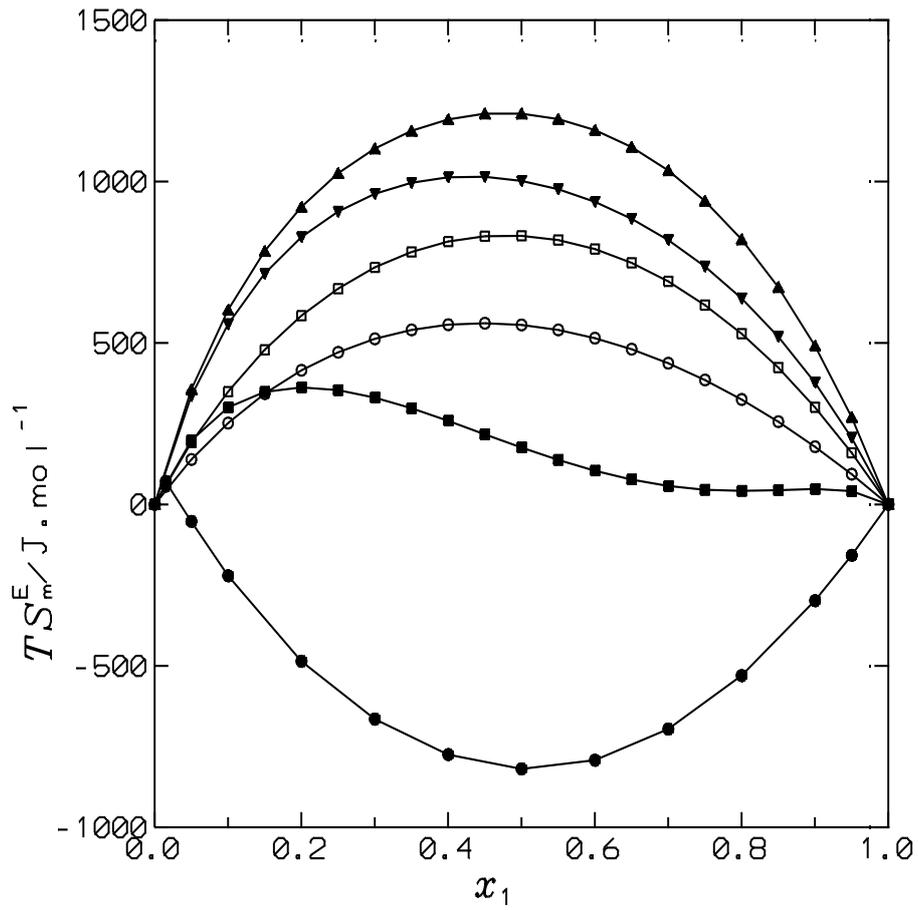

Figure 8. $TS_m^E$ curves for systems at 298.15 K and and 0.1 MPa: (●), ethanol(1) + hexane(2) [85]; (■), methanol(1) + DMC(2); (▲), 1-butanol(1) + DMC(2); (▼), 1-butanol(1) + DEC(2); (□) DMC(1) + heptane(2); (O), DEC(1) + heptane(2). Lines are only for the aid of the eye

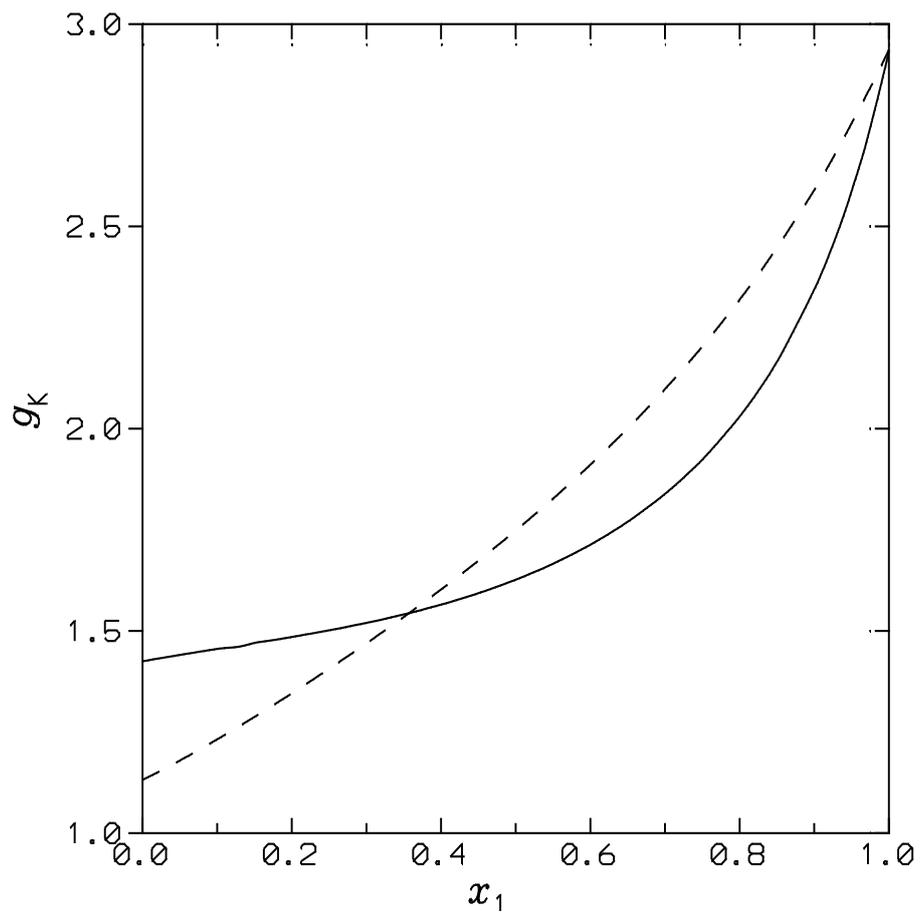

Figure 9. Kirkwood correlation factor, $g_K$, for the methanol(1) + ethylene carbonate(2) (solid line) [105], or + *N,N*-dimethylformamide (dashed line) [106,107] mixtures at 298.15 K and 0.1 MPa

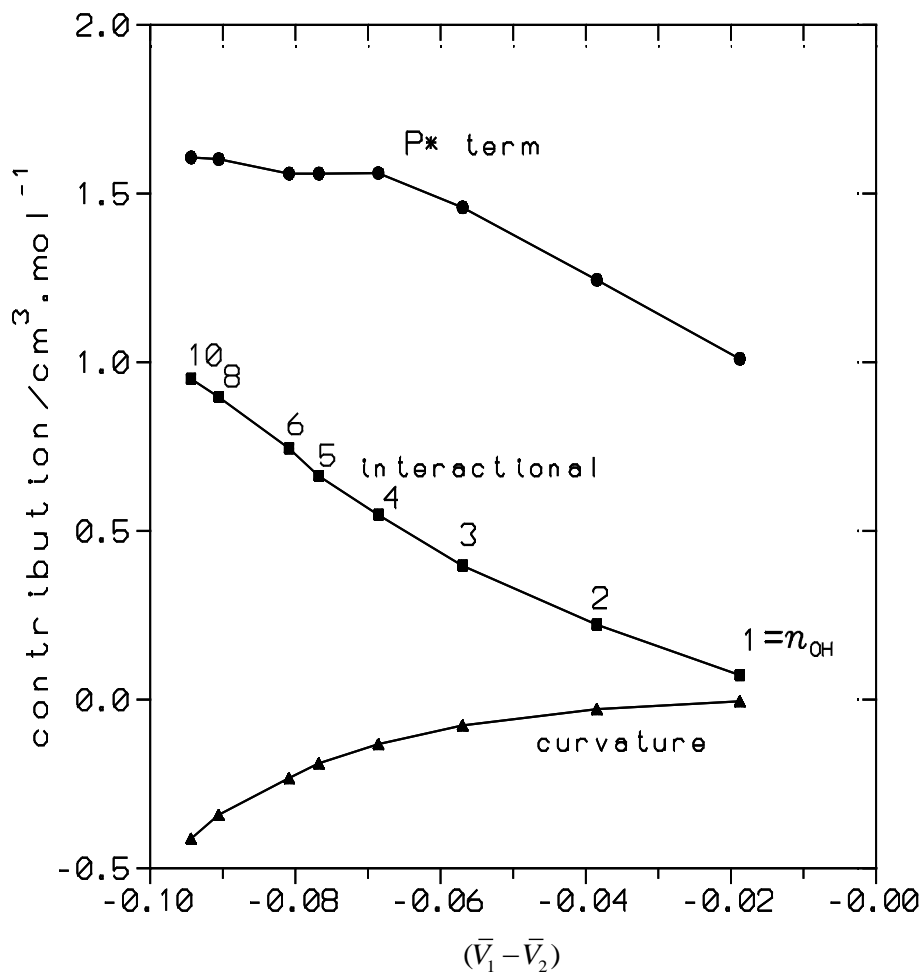

Figure 10. Interactional, curvature and $P^*$ contributions to $V_m^E$, calculated using the PFP model, for 1-alkanol(1) + DEC(2) mixtures at 298.15 K and equimolar composition vs. $(\bar{V}_1 - \bar{V}_2)$ the difference between the reduced volumes. $n_{OH}$ stands for the number of C atoms in the 1-alkanol. Lines are only for the aid of the eye.

# SUPPLEMENTARY MATERIAL

# ORIENTATIONAL EFFECTS IN MIXTURES OF ORGANIC CARBONATES WITH ALKANES OR 1-ALKANOLS

JUAN ANTONIO GONZÁLEZ[(1)*], FERNANDO HEVIA[(1)], CRISTINA ALONSO-TRISTÁN[(2)], ISAÍAS GARCÍA DE LA FUENTE[(1)] AND JOSE CARLOS COBOS[(1)]

[(1)] G.E.T.E.F., Departamento de Física Aplicada, Facultad de Ciencias, Universidad de Valladolid, Paseo de Belén, 7, 47011 Valladolid, Spain,

*e-mail: jagl@termo.uva.es; Fax: +34-983-423136; Tel: +34-983-423757

[(2)] Dpto. Ingeniería Electromecánica. Escuela Politécnica Superior. Avda. Cantabria s/n. 09006 Burgos, (Spain)

TABLE S1

Molar excess enthalpies, $H_m^E$, volumes, $V_m^E$, at 0.1 MPa, 298.15 K and equimolar composition for n-alkanone(1) + alkane(2) and for 1-alkanol(1) + n-alkanone(2) systems. Also included are the equation of state contribution ($\frac{\alpha_p}{\kappa_T} T V_m^E$) to the excess molar enthalpy and the excess molar internal energy at constant volume, $U_{Vm}^E$.

| Compound | $H_m^E$/J·mol$^{-1}$ | $V_m^E$/cm$^3$·mol$^{-1}$ | $\frac{\alpha_p}{\kappa_T} T V_m^E$/J·mol$^{-1}$ | $U_{Vm}^E$/J·mol$^{-1}$ |
|---|---|---|---|---|
| | 2-propanone (1) + n-alkane (2) | | | |
| Heptane | 1676 [S1] | 1.130 [S1] | 315 | 1361 |
| Decane | 1968 [S2] | 1.333 [S2] | 395 | 1573 |
| | 3-pentanone (1) + n-alkane (2) | | | |
| Heptane | 1078 [S3] | 0.520 [S4] | 145 | 933 |
| | 1-alkanol (1) + 2-propanone (2) | | | |
| Methanol | 686 [S5] | − 0.339 [S6] | − 106 | 792 |
| Ethanol | 1150 [S7] | − 0.071 [S6] | − 22 | 1172 |
| 1-butanol | 1537 [S8] | 0.050 [S9] | 16 | 1521 |
| | 1-alkanol (1) + 3-pentanone (2) | | | |
| Methanol | 725 [S10] | − 0.199 [S11] | − 63 | 788 |
| Ethanol | 987 [S10] | − 0.063 [S11] | − 20 | 1007 |
| 1-propanol | 1160 [S10] | − 0.036 [S11] | − 11 | 1171 |

TABLE S2

Contributions to $V_m^E$ according to the Prigogine-Flory-Patterson model for dialkyl carbonate + *n*-alkane, or 1-alkanol + organic carbonate systems at 298.15 K, 0.1 MPa and equimolar composition.

| System | $V_{m,int}^E$ / cm$^3$·mol$^{-1}$ | $P^*$ term/ cm$^3$·mol$^{-1}$ | Curvature term/ cm$^3$·mol$^{-1}$ | $V_m^E$ / cm$^3$·mol$^{-1}$ | Ref. |
|---|---|---|---|---|---|
| DMC + *n*-C$_7$ | 1.894 | − 0.013 | 0 | 1.158 | S12 |
| DMC + *n*-C$_{10}$ | 1.876 | 0.448 | − 0.061 | 1.442 | S12 |
| DEC + *n*-C$_7$ | 1.209 | 0.078 | − 0.002 | 0.7362 | S13 |
| DEC + *n*-C$_{10}$ | 1.288 | 0.604 | − 0.103 | 1.0629 | S13 |
| Methanol + DMC | 1.038 | 0.043 | − 0.002 | − 0.0628 | S14 |
| Ethanol + DMC | 1.385 | 0.172 | − 0.016 | 0.1635 | S15 |
| 1-propanol + DMC | 1.583 | 0.329 | − 0.052 | 0.3693 | S15 |
| 1-butanol + DMC | 1.900 | 0.459 | − 0.093 | 0.4771 | S16 |
| Methanol + DEC | 1.010 | 0.0723 | − 0.005 | − 0.048 | S17 |
| Ethanol + DEC | 1.244 | 0.222 | − 0.028 | 0.1143 | S17 |
| 1-propanol + DEC | 1.459 | 0.397 | − 0.078 | 0.2225 | S17 |
| 1-butanol + DEC | 1.561 | 0.548 | − 0.132 | 0.2815 | S17 |
| 1-hexanol + DEC | 1.559 | 0.745 | − 0.233 | 0.3940 | S17 |
| 1-octanol + DEC | 1.602 | 0.897 | − 0.342 | 0.5200 | S17 |
| 1-decanol + DEC | 1.607 | 0.951 | − 0.413 | 0.6386 | S17 |
| Methanol + PC | 0.745 | − 0.309 | − 0.071 | − 0.3362 | S18 |
| Ethanol + PC | 1.030 | − 0.323 | − 0.051 | − 0.2021 | S18 |
| 1-propanol + PC | 1.243 | − 0.251 | − 0.026 | − 0.037 | S18 |

**Literature cited (supplementary material)**